\documentclass[final,5p,times,twocolumn]{elsarticle}

\usepackage{amssymb}
\usepackage{amsmath}
\usepackage[T1]{fontenc}
\usepackage{blindtext}
\usepackage{slashed}
\usepackage{tikz-feynman}

\journal{Journal of Subatomic Physics ans Cosmology}

\begin{document}

\let\WriteBookmarks\relax
\def\floatpagepagefraction{1}
\def\textpagefraction{.001}

\begin{frontmatter}

\title{Speed of sound of QCD matter at chiral crossover}



\author{Oleksii Ivanytskyi$^a$}
\ead{oleksii.ivanytskyi@uwr.edu.pl}
\affiliation{organization={Incubator of Scientific Excellence—Centre for Simulations of Superdense Fluids,
University of Wroc\l aw},
            addressline={Max Born plac 9}, 
            city={Wroc\l aw},
            postcode={50-204}, 
            country={Poland}}

\author{David Blaschke$^{b,c,d}$}
\ead{david.blaschke@uwr.edu.pl}
\affiliation{organization={Institute of Theoretical Physics,
University of Wroc\l aw},
            addressline={Max Born plac 9}, 
            city={Wroc\l aw},
            postcode={50-204}, 
            country={Poland}}
\affiliation{organization={Helmholtz-Zentrum Dresden-Rossendorf (HZDR)},
            addressline={Bautzner Landstrasse 400}, 
            city={Dresden},
            postcode={01328}, 
            country={Germany}}
\affiliation{organization={Center for Advanced Systems Understanding (CASUS)},
            addressline={Untermarkt 20}, 
            city={G\"orlitz},
            postcode={02826}, 
            country={Germany}}

\author{Gerd R\"opke$^{e}$}
\ead{gerd.roepke@uni-rostock.de}
\affiliation{organization={Institute of Physics, University of Rostock},
            addressline={Albert-Einstein Str. 23-24}, 
            city={Rostok},
            postcode={D-18059}, 
            country={Germany}}  

\def\tsc#1{\csdef{#1}{\textsc{\lowercase{#1}}\xspace}}
\tsc{WGM}
\tsc{QE}


\begin{abstract}
Based on a generalized Beth-Uhlenbeck approach to thermodynamics of QCD motivated by cluster decomposition we present a unified equation of state of hot strongly interacting matter and analyze its properties in a wide range of temperatures.
The hadrons are treated as color singlet multiquark clusters in medium with a background gluon field in the Polyakov gauge. 
The confining aspect of QCD is accounted for by the Polyakov loop mechanism and by a large vacuum quark mass motivated by a confining density functional approach. 
We demonstrate that an abrupt switching between hadronic and partonic degrees of freedom, which is one of striking manifestations of dynamical restoration of chiral symmetry, is accompanied by a smooth behavior of entropy density at chiral crossover.
Individual contributions of different components of strongly interacting matter to its speed of sound are analyzed for the first time.
It is shown that restoration of chiral symmetry drives speed of sound of hadron gas to negative values, manifesting its mechanical instability and being in a strike disagreements with the lattice QCD data.
Accounting for the partonic excitations naturally resolves this contradiction.
\end{abstract}





\end{frontmatter}


\section{Introduction}
\label{intro}

Extreme complexity of Quantum Chromodynamics (QCD) significantly obscures probing its phase diagram, which has been in the focus of intense theoretical \citet{Fukushima:2010bq}, experimental \citet{Busza:2018rrf} and astrophysical studies \citet{Baym:2017whm} during the last decades.
Dynamical restoration of chiral symmetry and deconfinement of color non-singlet quarks and gluons, which are also referred to as partons, are among the key topics related the QCD phase diagram.

Currently, only Monte–Carlo simulations of the QCD partition function on discretized Euclidean space–time lattices (see e.g. \citet{Borsanyi:2010cj,Borsanyi:2013bia,HotQCD:2014kol,HotQCD:2018pds,Borsanyi:2021sxv}) allow addressing these questions from the first principles and directly probing equation of state (EoS) of QCD.
However, even the lattice QCD simulations are limited to the region of the phase diagram, where baryon chemical potential does not exceed three times the temperature. 
At these regimes the QCD phase diagram is not expected to contain any phase transitions manifested by an irregular behavior of some thermodynamic quantities, which can be relatively easily detected and used as an indication of restoration of chiral symmetry and/or deconfinement.
Contrary, a smooth crossover between hadrons and partons is expected at small baryonic chemical potentials.
Thus, detecting both restoration of chiral symmetry and deconfinement of color using just bulk thermodynamic data seems to be problematic.
This explains a need of information about composition of QCD matter.

This information can be obtained only within a unified approach, which simultaneously contains hadronic and partonic degrees of freedom.
At the same time, suppression of color non-singlet quarks and gluons at small temperatures as well as dissociation of hadrons at high temperatures should be accounted for within such approach to provide its qualitative agreement with the expected phenomenology of the QCD phase diagram.
Suppression of partons at small temperatures can be provided by large quark selfenergies motivated by the confining density functional developed in Refs. \citet{Ivanytskyi:2022bjc,Ivanytskyi:2022oxv} and the Polyakov loop mechanism (see e.g. Refs. \cite{Ratti:2006wg,Lo:2013hla}).
Allowing the hadron dissociation at high temperatures requires treating them not as elementary quasiparticles, but as effective in-medium correlations of (anti)quarks.
This task can be systematically resolved within the approach of cluster decomposition, which was proposed for hot fermionic systems with clusterization by \citet{Baym:1961zz,Baym:1962sx}.

In this work we apply a unified hadron-parton EoS earlier developed in Ref. \cite{Blaschke:2023pqd} and motivated by the cluster decomposition of QCD matter to study its thermodynamics and composition in the region of chiral crossover.
We for the first time analyze individual contributions of various components of QCD matter to the stiffness of the QCD EoS and inspect the question of their mechanical stability.
For this we calculate and confront to the data of lattice QCD total speed of sound of QCD matter as well as individual speeds of sound of its components.
We conclude that mechanical stability of QCD matter requires excitation of partonic degrees of freedom immediately after the pseudocritical temperature of chiral crossover.  
This implies a sharp switching between hadrons and partons, which cannot be detected by a smooth behavior of various thermodynamic quantities.

The paper is organized as follows. 
In the next section we formulate the model motivated by the cluster decomposition for QCD matter.
Its entropy density and speed of sound are analyzed and discussed in Section \ref{speed}.
The conclusions are given in Section \ref{concl}.

\section{Cluster decomposition for QCD matter}
\label{cluster}

The so-called $\Phi$-functional approach by \citet{Baym:1961zz,Baym:1962sx} allows systematically treating many particle  correlations in thermal medium, which is equivalent to cluster decomposition for QCD matter.
Within this scheme quarks appear as monomers, while hadrons and multiquark colored states, e.g. diqurks, are treated on common footing as clusters of different sizes.
The Green's functions of all the clusters including monomers should be simultaneously found as a solution a system of coupled Schwinger-Dyson equations with the cluster self-energies defined as functional derivatives of the $\Phi$-functional with respect to the corresponding propagators.
Restricting the $\Phi$-functional to the class of two-loop irreducible diagrams of the sunset type is equivalent to the generalized Beth-Uhlenbeck approach to the problem of clusterizaton in hot and/or dense fermionic systems \citet{Ropke:2012qv}.
The further simplification is related to treating quarks at the mean-field level, which allows subsequent determination of the Green's functions of clusters $\hat S_n$ of all possible sizes $n$ using the Green's functions of clusters of smaller sizes $\hat S_{m<n}^{-1}$.
This scheme motivated formulation of a unified model of clusterized QCD matter from Ref. \cite{Blaschke:2023pqd}.
In this section we briefly introduce the model.
For simplicity we consider the case of zero baryon density, when temperature $T$ is the only thermodynamic parameter.

The chiral dynamics of quarks is modeled by an abrupt switching of the quark masses $M_f$ from the vacuum $M_{u,d}|_{T=0}=627$ MeV, $M_s|_{T=0}=770$ MeV to the current $m_{u,d}=5.6$ MeV, $m_s=124$ MeV values at the pseudocritical temperature $T_c=156.6$ MeV, so that
\begin{eqnarray}
    \label{I}
    M_f=\theta(T_c-T)M_f|_{T=0}+\theta(T-T_c)m_f,
\end{eqnarray}
where the subscript index ``$f$'' labels quark flavors.
The vacuum quark masses were fixed according to the procedure described below, while the current masses were chosen according to the Review of Particle Physics by \citet{ParticleDataGroup:2022pth}.
It is worth mentioning that such high quark masses in the region of broken chiral symmetry are motivated by the confining density functional approach to quark matter developed by \citet{Ivanytskyi:2022oxv,Ivanytskyi:2022bjc}.
A more elaborate modeling of chiral dynamics would correspond to determining the effective quark masses within a self-consistent chiral quark model.
This task has been successfully accomplished within the mentioned chiral confining density functional approach to quark matter.
For the sake of simplicity, in this work we omit this step and model spontaneous breaking of chiral symmetry and its dynamical restoration by simply prescribing the temperature dependence of the quark masses, which treats $T_c$ as an input parameter.
Its used value is fixed according to the lattice QCD data on the temperature of chiral crossover \cite{HotQCD:2018pds}.

The described chiral dynamics is gauged through the running coupling of QCD $g$ by the Polyakov gluon field $A^\mu=A_a^\mu\lambda_a$, where $\lambda_a$ are the SU$(N_c)$ color group generators at $N_c=3$.
The gluon field enters the in-medium quark propagators as $\hat S_1^{-1}=i\slashed\partial+g\slashed A-\hat M$ and is represented by the Polyakov loop potential $U_\phi$.
The latter explicitly depends on temperature, the Polyakov loop variable $\phi={\rm tr}_c\exp(gA_3^0\lambda_3/T)/N_c$ and its complex conjugate.
The trace in definition of $\phi$ is carried over the color indexes.
In this work we use the potential $U_\phi$ from Ref. \citet{Lo:2013hla}.
Note, at zero baryon density considered in this work the Polyakov loop variable is real, so below we use $\phi^*=\phi$.

The contribution of perturbative quark states with three-momenta $|{\bf k}|>\Lambda_{\rm pert}=222$ MeV is accounted for by the order $\mathcal{O}(\alpha_s)$ perturbative correction of massless quarks gauged by the Polyakov gluon field.
Running of the QCD coupling leads to
\begin{eqnarray}
    \label{II}
    \alpha_s=\frac{g^2}{4\pi}=
             \frac{12\pi}{11N_c-2N_f}
             \left[\ln^{-1}\frac{T^2}{T_*^2}-\frac{T_*^2}{T^2-T_*^2}\right],
\end{eqnarray}
where $N_f=3$ is number of quark flavors and $T_*=94$ MeV.
The Landau pole in this expression is removed according to the procedure of Analytic Perturbation Theory proposed by \citet{Shirkov:2001sm}.

Thermodynamic potential of a hadron or a colored multiquark cluster of sort $i$ composed by $n_i$ (anti)quarks is given by the generalized Beth-Uhlenbeck formula
\begin{eqnarray}
    \Omega_i=(-1)^{n_i}d_iT&&\hspace*{-.7cm}
    \int\frac{d{\bf k}}{(2\pi)^3}
    \\\nonumber
    \label{III}
    &&\hspace*{-.7cm}
    \int\frac{d\omega}{\pi}\ln\left(1+(-1)^{n_i+1}f_i\right)
    \sin^2\delta_i\frac{\partial\delta_i}{\partial\omega},
\end{eqnarray}
where $d_i$ is  spin-isopin-color  degeneracy factor of this cluster.
Note, the color factor in the degeneracy factors of color-singlet hadrons is one.
The thermal distribution function in Eq. (\ref{III}) reads
\begin{eqnarray}
    \label{IV}
    \hspace*{-.7cm}
    f_i=\frac{\partial}{\partial\ln\omega}
         \left\{\hspace*{-.2cm}
         \begin{array}{l}
              \ln\left(1+y_i\right)
              \hspace*{2.8cm}{\rm hadrons}\\
              \ln\left(1+3\phi y_i+3\phi y_i^2+y^3\right)
              N_c^{-1}
              ~{\rm colored~clusters}
         \end{array}     
         \right.\hspace*{-.2cm},\hspace*{-.4cm}
\end{eqnarray}
where $y_i=(-1)^{n_i+1}e^{\omega/T}$.
As is seen from this expression, colored multiquark states, e.g. diquarks, are gauged by the Polyakov gluon field as well as quarks.
The thermal distributions of colored multiquark clusters include the factor $N_c^{-1}$, which compensates the double counting of their color states already accounted for in the degeneracy factors $d_i$.
The need of this factor can be seen by setting the Polyakov loop variable $\phi=1$ and comparing the distributions of hadrons and colored multiquark clusters. 

The spectral properties of multiquark clusters are carried by the phase shifts $\delta_i=\Im\ln\hat S_i$.
Defining the latter requires the Green's functions of multiquark clusters, which can be constructed using the corresponding self-energies.
However, to facilitate calculations in Ref. \citet{Blaschke:2023pqd} these phase shifts were defined via the parametric model.
The corresponding phase shifts in the rest-frame of a multiquark cluster (${\bf k}=0$ and $\omega=M$) are 
\begin{eqnarray}
    \hspace*{-.6cm}
    \delta_i
    =\delta^*_i \Biggl[\theta(M_{{\rm thr},i}-M)
    \hspace*{-.2cm}&+&\hspace*{-.2cm}
    \frac{1}{\pi}{\rm acos}\left(\frac{2 M-2 M_{{\rm thr},i}-N_i\Lambda_i}{N_i\Lambda_i}\right)
    \nonumber\\
    \label{V}
    \hspace*{-.2cm}&\times&\hspace*{-.2cm}
    \theta(M-M_{{\rm thr},i}) \theta(M_{{\rm thr},i}+\Lambda_i N_i-M)\Biggl]
\end{eqnarray}
with 
\begin{equation}
\label{VI}
    \delta^*_i(M)
    =
    \begin{cases}
        \pi\, \theta(M-M_i)
       ~,&T<T_c\\
        \pi\theta(M-M_{{\rm thr},i})
        \frac{\mathrm{atan}\left(\frac{M^2-M_i^2}{\Gamma_iM_i}\right)-\mathrm{atan}\left(\frac{ M_{{\rm thr},i}^2-M_i^2}{\Gamma_iM_i}\right)}
        {\frac{\pi}{2}-\mathrm{atan}\left(\frac{ M_{{\rm thr},i}^2-M_i^2}{\Gamma_iM_i}\right)}~.&T>T_c
    \end{cases}.
\end{equation}
Eqs. (\ref{V}) and (\ref{VI}) include $N_i=N_{l,i}+N_{s,i}$ with $N_{l,i}$ and $N_{s,i}$ being the numbers of light and strange (anti)quarks constituting the multiquark cluster of sort $i$.
The mass and decay width of this cluster are defined as
\begin{eqnarray}
    \label{VII}
    M_i&=&M_{i,T=0}+\alpha_1\, \Gamma_i,\\
    \label{VIII}
    \Gamma_i&=&\alpha_2\, (T-T_c)\theta(T-T_c),
\end{eqnarray}
respectively, while $\alpha_1=11.4$ and $\alpha_2=1.9$ are adjusted in agreement with the lattice data on entropy density.
The vacuum masses of hadrons are fixed according to the Review of Particle Physics by \citet{ParticleDataGroup:2022pth}.
Since colored multiquark clusters have not been observed experimentally, their vacuum masses remain unknown.
In this work we estimate them as
\begin{eqnarray}
    \label{IX}
    M_{i,T=0}=N_{l,i}M_{l,T=0}+N_{s,i}M_{s,T=0}-(N_i-1)B,
\end{eqnarray}
where $B=471$ MeV is the binding energy per quark bond.
This binding energy was determined along with the vacuum quark masses by fitting Eq. (\ref{IX}) to the masses of the ground states of nucleons, pseudoscalar and vector mesons (see Ref. \citet{Blaschke:2023pqd} for detail).
The temperature dependent decay thresholds of multiquark clusters in Eqs. (\ref{V}) and (\ref{VI}) are
\begin{eqnarray}
    \label{X}
    M_{{\rm thr},i}=N_{l,i}M_l+N_{s,i}M_s.
\end{eqnarray}
\begin{figure}[!htb]
\centering
\includegraphics[width=\columnwidth]{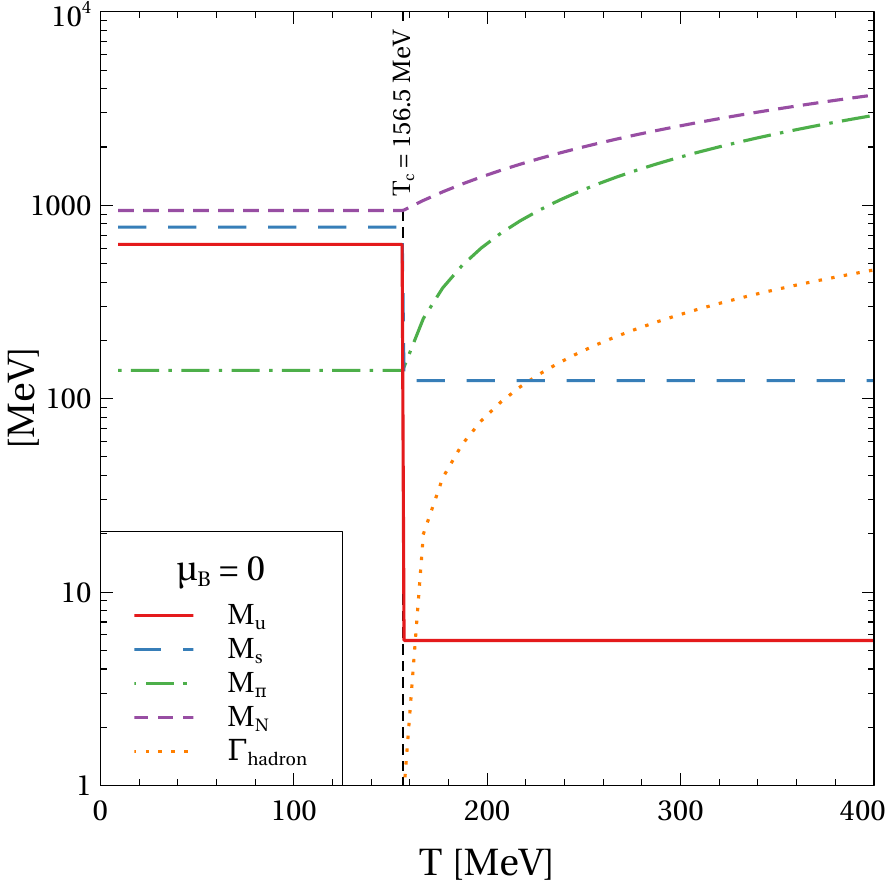}\\
\includegraphics[width=\columnwidth]{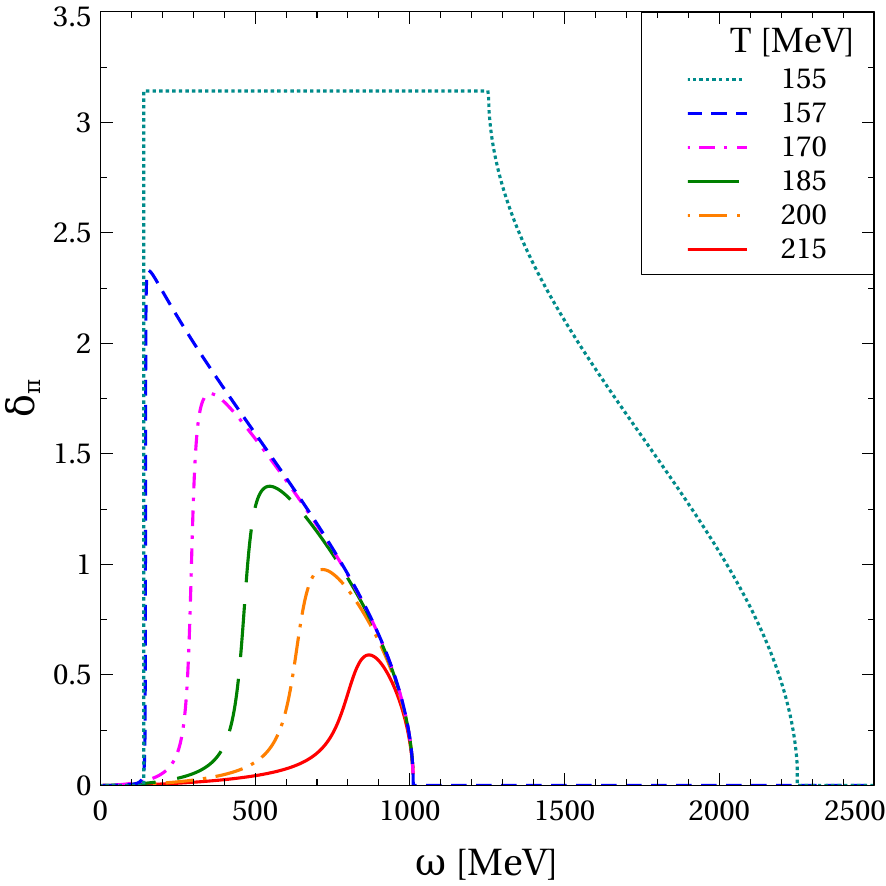}
\caption{Temperature dependence of the mass spectrum of quarks, pions and nucleons as well as the decay width of the Breit-Wigner model of these hadrons (top panel) and parametric phase shift of pions in the rest-frame used as a generic example (bottom panel).
The dashed line in the top panel indicates the temperature of chiral cross over.
The numbers in the legend in the bottom panel correspond to the temperature at which the phase shift was calculated.
The calculations are performed at vanishing chemical potential.}
\label{fig1}
\end{figure}

The described scheme of treating multiquark clusters corresponds to the Mott Hadron Resonance Gas (MHRG) picture of the low temperature phase of QCD.
As is seen from Fig. \ref{fig1}, below the pseudocritical temperature quarks are heavy and masses of hadrons and colored multiquark clusters attain their vacuum values.
This provides $M_{{\rm thr},i}>M_i$, which makes quark decays of clusters kinematically forbidden.
Thus, at $T<T_c$ hadrons and colored multiquark clusters are stable bound states with vanishing widths.
At the same time, the latter are suppressed by the Polyakov loop mechanism.
At the pseudocritical temperature and above it quarks become light, which makes quark decays of multiquark clusters  possible.
This manifests Mott dissociation of clusters.
At $T>T_c$ this clusters exist as short-living resonances with finite width.
The parameterization (\ref{VII}) and (\ref{VIII}) is motivated by the results extracted from the analysis of the polarization operator of pions in the two-loop approximation by \citet{Blaschke:2013zaa}.

The MHRG picture is also reflected in the behavior of the rest-frame phase shifts shown in Fig. \ref{fig1}.
The bound state nature of hadrons and colored multiquark clusters below the pseudocritical temperature is accounted for by the step up of $\delta_i$ at $\omega=M_i$, which leads to infinitely thin positive contribution to the spectral function $\sin^2\delta_i\partial\delta_i/\partial\omega=\delta(\omega-M_i)$.
Smooth decreasing part of $\delta_i$ at $\omega>M_{{\rm thr},i}$ corresponds to quark continuum of hadrons and colored multiquark clusters.
Its produces a negative contribution in their spectral function. 
The described elements of the phase shift are separated by its flat part with $\delta_i=\pi$ at $\omega\in[M_i,M_{{\rm thr},i}]$.
It disappears at the Mott temperature, which coincides with the pseudocritical one, since $M_i=M_{{\rm thr},i}$ in this case.
Above this temperature masses of hadrons and colored multiquark clusters exceed their decay thresholds and the widths are finite.
This results in a smooth phase shift of the resonant type  without a flat part of heights $\pi$.

The total thermodynamic potential of the present model can be written as
\begin{eqnarray}
   \label{XI}
   \Omega=-\frac{T}{V}{\rm tr}\ln\left(\hat S_1^{-1}\right)+\mathcal{U}_\phi+\sum_i\Omega_i+\frac{T}{2V}
\begin{array}{c}
\scalebox{.6}{
\begin{tikzpicture}
\begin{feynman}
\vertex [dot] (a) {};
\vertex [right = of a, dot]  (b) {};
\diagram{(b) -- [fermion,half right] (a)};
\diagram{(a) -- [gluon] (b)};
\diagram{(a) -- [fermion,half right] (b)};
\end{feynman}
\end{tikzpicture}}
\end{array},
\end{eqnarray}
where $V$ is spacial volume of the system and the last term stands for the two-loop perturbative correction
\begin{eqnarray}
   \label{XII}
   \hspace*{-.8cm}
   \frac{T}{2V}
   \hspace*{-.1cm}
\begin{array}{c}
\scalebox{.6}{
\begin{tikzpicture}
\begin{feynman}
\vertex [dot] (a) {};
\vertex [right = of a, dot]  (b) {};
\diagram{(b) -- [fermion,half right] (a)};
\diagram{(a) -- [gluon] (b)};
\diagram{(a) -- [fermion,half right] (b)};
\end{feynman}
\end{tikzpicture}}
\end{array}
\hspace*{-.2cm}=\frac{32\pi\alpha_s}{3}
\left[\frac{T^2}{2}\int\frac{d{\bf k}}{(2\pi)^3}\frac{f_1}{|{\bf k}|}+
\left(\int\frac{d{\bf k}}{(2\pi)^3}\frac{f_1}{|{\bf k}|}\right)^2\right].
\end{eqnarray}
As is seen, the Feynman graph in Eq. (\ref{XII}) includes both quark and gluon lines.
Thus, the order $\mathcal{O}(\alpha_s)$ correction accounts for the contributions of not only perturbative quark states but also gluons mediating scattering processes among them. 
Note, the purely gluon order $\mathcal{O}(\alpha_s)$ graph, which can be obtained from the graph in Eqs. (11) and (12) by replacing the quark lines by the gluon ones, is absent in the perturbative correction to the thermodynamic potential.
This is because all the purely gluonic contributions, both perturbative and non-perturbative, are already accounted for in the Polyakov loop potential, which is fitted to the lattice data of pure gauge QCD. 

The Polyakov loop variable is defined by minimizing the thermodynamic potential, i.e. as a solution of the equation $\partial\Omega/\partial\phi=0$. 
Having it found, the entropy density and squared speed of sound can be obtained using the well known thermodynamic identity $s=-\partial\Omega/\partial T$ and the relation $c_S^2=\partial\ln T/\partial\ln s$.
This expressions can be applied to define partial contributions of different components to the entropy density as $s_a=-\partial\Omega_a/\partial T$ ($a$= (non)perturbative quarks, gluons, hadrons and colored multiquark clusters) as well as their individual speeds of sound $c_{Sa}^2=\partial\ln T/\partial\ln s_a$.
For non-perturbative quarks $s_a$ is calculated as the entropy density of fermion quasiparticles with the mass (\ref{I}), which are gauged by the Polyakov gluon field.
The entropy density of gluons and perturbative quarks are directly found by differentiating the Polyakov loop potential and Eq. (\ref{XII}), respectively.
The entropy density of hadrons and colored multiquark clusters should be evaluated as partial $T$-derivatives of their thermodynamic potentials $\Omega_i$.
It is important to mention, that phase shifts entering these $\Omega_i$ also depend on temperature. 
At the same time, the terms arising from their differentiation are exactly canceled by the corresponding terms of the temperature derivative of $\Phi$-functional if the latter is limited to the two-loop skeleton
diagrams \cite{Vanderheyden:1998ph,Blaizot:2000fc}.
Explicit expressions of $s_a$ can be found in Ref. \cite{Blaschke:2023pqd}.

Before analyzing the properties of QCD matter at chiral crossover, we would like to discuss the physical meaning of the individual speeds of sound.
Customizing their definition as $ds_a/dT=s_a/Tc_{S,a}^2$ and using the second law of thermodynamics $dQ_a=T ds_a$ with the amount of transferred heat expressed through the corresponding individual heat capacities as $dQ_a=C_a dT$, the later can be written as
\begin{eqnarray}
    \label{XIII}
    C_a=\frac{s_a}{c_{S,a}^2}.
\end{eqnarray}
From this relation we can conclude that negative $c_{s,a}^2$, observed for hadrons and colored multiquark clusters above the pseudocritical temperature, lead to the instability of thermal equilibrium signaled by a negative heat capacity unless the total speed of sound and, consequently, heat capacity is positive due to the onset of partons. 
Thus, individual speeds of sounds allow probing stability of thermal equilibrium of individual components of QCD matter.

In Ref. \citet{Blaschke:2023pqd} and in this study we account for 58 mesonic, 60 baryonic, 60 antibaryonic hadrons with the masses below 2.6 GeV as well as deutron and hypothetical spin-flavor-color singlet sexaquark state described in Ref. \citet{Buccella:2020mxi} along with their antiparticles. 
Diquarks, tetraquarks and pentaquarks of all the possible flavor contents also are accounted for as well as their antiparticles.

\section{Speed of sound at chiral crossover}
\label{speed}

\begin{figure}[!t]
\centering
\includegraphics[width=\columnwidth]{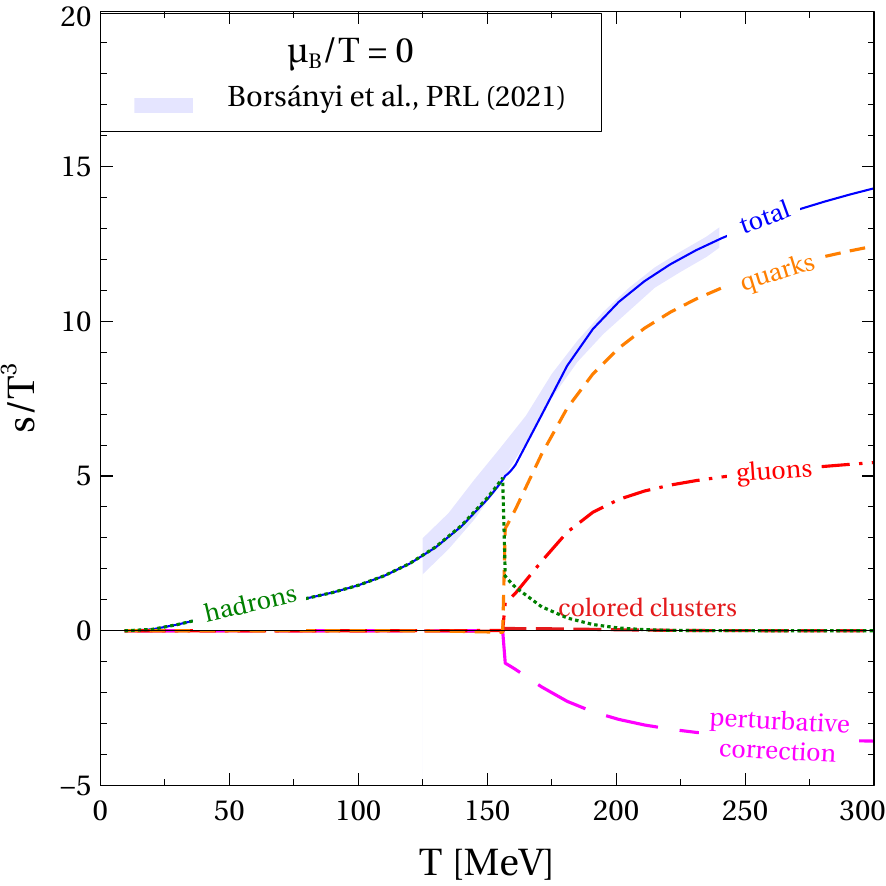}\\
\includegraphics[width=\columnwidth]{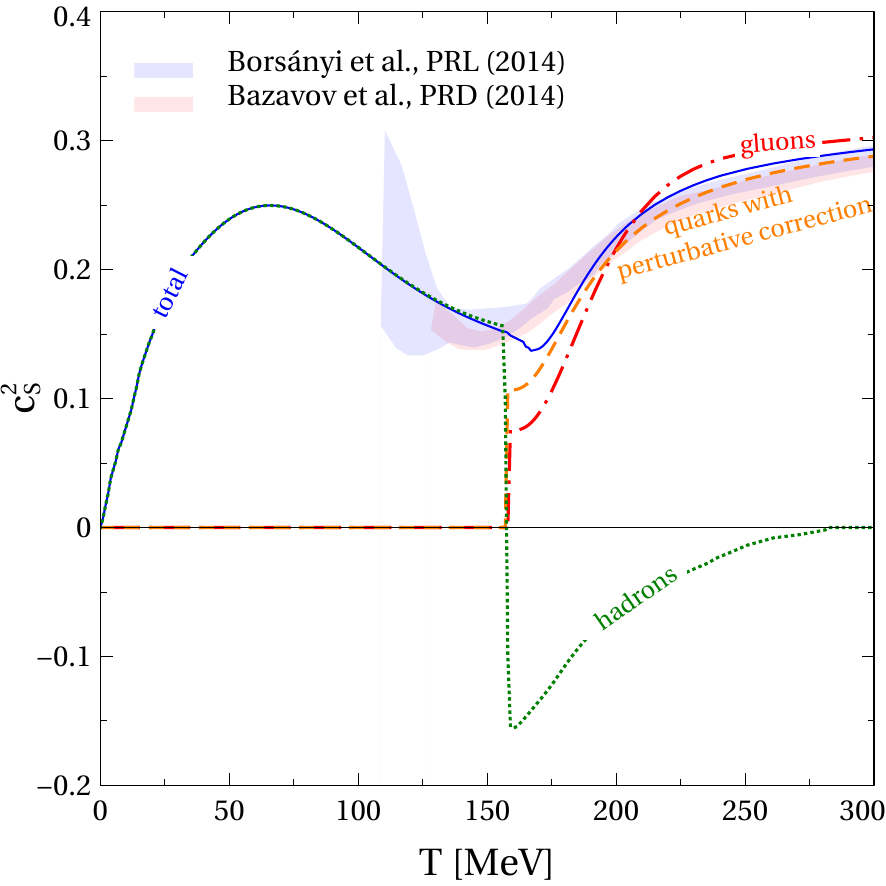}
\caption{Temperature dependence of entropy density (top panel) and speed of sound (bottom panel) of QCD matter.
The shaded areas represents the lattice QCD data from Refs. \cite{Borsanyi:2021sxv,Borsanyi:2013bia,HotQCD:2014kol}.
The calculations are performed for vanishing chemical potential}
\label{fig2}
\end{figure}

Defining temperature dependence of the entropy density of QCD matter is the first step toward determining its speed of sound.
Fig. \ref{fig2} shows $s$ introduced above as a function of $T$. 
Below the pseudocritical temperature the entropy density is dominated by the MHRG of weakly interacting hadrons, which can be rather accurately described by the Hadron Resonance Gas model, assuming that hadron interactions are saturated to formation of resonances \citet{Hagedorn:1965st}.
The contributions of quarks, gluons and colored multiquark clusters are suppressed at this domain since they are color non-singlet states. 
At $T=T_c$ hadrons and colored multiquark clusters experience the Mott dissociation, above which they exists as short living resonances.
This diminishes their contribution to the entropy density.
At the same time, hadron dissociation deconfines quarks.
Their thermal excitations produce the main contribution to the entropy density above the pseudocritical temperature.
Being gauged by the gluon field, quark excitations also drive growth of the Polyakov loop variable at $T>T_c$.
This manifests thermal excitation of gluons, which also valuably contribute to $s$.
At the same time, colored multiquark clusters remain suppressed even at $T>T_c$ due to their Mott dissociation to the constituent quarks.
However, the colored clusters are characterized by large baryon charges and spin-isospin degeneracy factors.
Due to this they carry significant fraction of the baryon charge and, therefore, play a significant role at finite baryon chemical potentials \cite{Blaschke:2023pqd}.
For example, at $\mu_B/T=3$ the partial entropy carried by the colored multiquark clusters is compatible to the one carried by gluons.
It is worth mentioning that the model also probes the thermodynamic importance of yet unobserved sexaquarks, which become abundant at finite baryon densities.
As expected, the contribution of quarks asymptotically saturates to the Stefan-Boltzmann limit, while accounting for the perturbative correction provides its agreement with the $\mathcal{O}(\alpha_s)$ perturbative results.

As is seen from Fig. \ref{fig2}, switching between colorless hadronic and colored partonic degrees of freedom at the pseudocritical temperature is abrupt and sharp in the sense that partial contributions of different components of QCD matter are discontinuous at $T_c$.
Nevertheless, behavior of the total entropy density is continuous and smooth and agrees well with the lattice QCD data from Ref. \citet{Borsanyi:2021sxv}.
This regularity of $s$ and its temperature derivative along with the sharp switching between quarks and hadrons is also consistent with the physical picture of quark-hadron duality \cite{Poggio:1975af}.
Technically, the regularity of $s$ is provided by the perfect cancellation of the amplitudes of drops and jumps of partial contributions of hadrons, multiquark clusters, quark and gluons to the total entropy density at the pseudocritical temperature.
The corresponding speed of sound also provides a reasonable agreement with the lattice QCD data from Refs. \citet{Borsanyi:2013bia,HotQCD:2014kol}.

It is worth mentioning that chiral condensate evaluated within the present approach is also continuous and smooth at the pseudocritical temperature even despite the fact that quark mass is discontinuous at $T_c$. 
Similarly to the entropy density, the discontinuity of the quark contribution to the total chiral condensate is compensated by the discontinuity of hadron chiral condensate, which can be evaluated within the formalism of cluster decomposition using hadronic $\sigma$-terms \cite{Jankowski:2012ms}.

Below the pseudocritical temperature thermodynamic properties of QCD matter are mostly defined by hadrons and their individual speed of sound coincides with the total one.
At low temperatures the total entropy density attains its maximum value at $T\simeq60$ MeV.
Above this temperature the hadron states heavier than pions get excited, which lowers the individual speed of sound of the MHRG.
The individual speeds of sound of quarks and gluons vanish at $T<T_c$ since partons are confined at low temperatures.
Above the pseudocritical temperature hadrons dissociate and the MHRG becomes mechanically unstable, which is seen from the negative value of its individual speed of sound.
Moreover, thermal equilibrium of the MHRG is also unstable in this case.
Thus, mechanical stability of thermally equlibrated QCD matter requires exciting partonic degrees of freedom immediately after $T_c$.
Fig. \ref{fig2} demonstrates that individual speeds of sound of quarks and gluons rapidly grow above the pseudocritical temperature and asymptotically saturate to the conformal value $c_S^2=1/3$. 
This stabilizes QCD matter at high temperatures. 
The total speed of sound has a minimum at a temperature slightly above the pseudocritical one.
This reflects softening of QCD matter in the vicinity of chiral cross over due to the deconfinement of partonic degrees of freedom.

It is also interesting that at the individual speed of sound of quarks exceeds the one of gluons right above the pseudocritical temperatures, where the total entropy density of nonperturbative and perturbative quark states also is above the entropy density of gluons.
This means that at the mentioned temperatures stiffness of QCD matter is mostly caused by quarks.
At higher temperatures, the individual speed of sound of gluons is higher than the one of quarks.
At the same time, as is seen from the top panel of Fig. \ref{fig2}, the entropy density carried by nonperturbative and perturbative quark states exceeds the entropy density carried by gluons at any temperature.
Consequently, at high temperatures quarks and gluons contribute to the stiffness of QCD matter compatibly.

\section{Conclusions}
\label{concl}

A unified EoS, that accounts for hadrons, quarks, diquarks, tetraquarks, pentaquarks and gluons, is formulated based on the generalized Beth-Uhlenbeck approach to hot fermionic systems with clusterization.
The EoS is applied to modeling QCD matter at zero baryon density.
The proper chiral dynamics of quark matter is modeled by a sharp switching of the quark masses from the high vacuum values to the relatively small values of the current quark masses.
The confining aspect of QCD is introduced to the model by gauging the colored quarks and multiquark states by the Polyakov gluon field.

The model is applied to analyze not only the bulk thermodynamic properties of hot QCD matter but also its composition.
A special attention is payed to the region of chiral cross over near the pseudocritical temperature.
While partial contributions of hadrons, quarks, colored multiquark states and gluons exhibit a dramatic irregularities at the pseudocritical temperature, the total entropy density remains smooth in agreement with the data of lattice QCD, which agrees with the crossover nature of the QCD transition at zero baryon density.
The contribution of colored multiquark clusters is found to be suppressed by the Polyakov loop mechanism at low temperatures and by their Mott dissociation at high temperatures. 

The speed of sound of clusterized QCD matter is analyzed for the first time.
It is shown that restoration of chiral symmetry causes mechanical instability of the MHRG manifested by negative values of its individual speed of sound above the pseudocritical temperature. 
This implies that thermodynamic properties of QCD matter are determined by partonic degrees of freedom immediately after chiral cross over.
Accounting for them provides reasonable consistency of the present model with the lattice QCD data on speed of sound.

While the general character of the approach allows its application at arbitrary baryon densities, reliable modeling of the QCD phase diagram at large $\mu_B$ should be performed with accounting for the possibility of existence of a first order phase transition of QCD and its critical endpoint.
This task can be accomplished within the chirally symmetric confining density functional approach \cite{Ivanytskyi:2022bjc,Ivanytskyi:2022oxv}, which is beyond the scope of the present work.

\vspace{0.5cm}
\section*{Acknowledgments}
The work of O.I. was performed within the program Excellence Initiative--Research University of the University of Wrocław of the Ministry of Education and Science.
O.I. and D.B. acknowledge support from the Polish National Science Center under grant No. 2021/43/P/ST2/03319.
The work of G.R. received support via a joint stipend from the Alexander von Humboldt Foundation and the Foundation for Polish Science.

\bibliographystyle{elsarticle-harv} 
\bibliography{refs}

\end{document}